\documentclass[submission]{eptcs}

\providecommand{\event}{ICLP DC 2020} 

\usepackage{breakurl}             
\usepackage{underscore}           
\usepackage{url}

\usepackage{xcolor}

\usepackage{listings}
\lstdefinelanguage{haskell}{
  keywords={data,newtype,type,let,in,where,case,of,if,then,else,class,instance,deriving,module,do}
}
\lstnewenvironment{haskell}{\lstset{language=haskell}}{}
\lstdefinelanguage{curry}{
  language=haskell,
  morekeywords={free}
}
\lstnewenvironment{curry}{\lstset{language=curry}}{}
\newcommand{\inv}{\ensuremath{^{\texttt{\scriptsize -1}}}}

\title{Research Summary on\\Implementing Functional Patterns by\\Synthesizing Inverse Functions}
\author{Finn Teegen
\institute{Kiel University\\
Kiel, Germany}
\email{fte@informatik.uni-kiel.de}
}
\def\titlerunning{Implementing Functional Patterns by Synthesizing Inverse Functions}
\def\authorrunning{F. Teegen}
\begin{document}
\maketitle

\begin{abstract}
In this research summary we present our recent work on implementing functional patterns with inverse functions in the lazy functional-logic programming language Curry.
Our goal is the synthesis of the inverse of any given function in Curry itself.
The setting of a functional-logic language especially allows the inversion of non-injective functions.
In general, inverse computation is a non-trivial problem in lazy programming languages due to their non-strict semantics.
We are so far able to directly derive the inverse function for a limited class of functions, namely those consisting of rules that do not involve both extra variables and non-linear right-hand sides.
Because the synthesized definitions are based on standard code, known optimizations techniques can be applied to them.
For all other functions we can still provide an inverse function by using non-strict unification.
\end{abstract}

\section{Introduction}
\label{sec:introduction}

Curry \cite{Hanus2016CurryReport} is a functional-logic programming language \cite{AntoyHanus2010FunctionalLogicProgramming} with Haskell-like syntax and support for non-deterministic operations \cite{GonzalezEtAl1999Approach} and free variables.
In standard code only variables and constructors are allowed in patterns on the left-hand side of function definitions.
Functional patterns \cite{AntoyHanus2005FunctionalPatterns} are an extension to Curry and relax this restriction by additionally allowing the use of function symbols within patterns.
As an example, we consider the function that returns the last element of a list.
\begin{curry}
last :: [a] -> a
last (xs++[x]) = x
\end{curry}
Here, we use the function \lstinline|(++)|, which concatenates two lists, to specify that the variable \lstinline|x| should correspond to the last element of the input list.
Semantically, the functional pattern is equivalent to the following infinite set of rules.
\begin{curry}
last [x]       = x
last [x1,x]    = x
last [x1,x2,x] = x
§$\ldots$§
\end{curry}
Note that the definition of \lstinline|last| using functional patterns is less strict than the following common encoding with an equality constraint, where \lstinline|(=:=)| denotes the (strict) unification operator.
\begin{curry}
last z | xs ++ [x] =:= z = x where xs, x free
\end{curry}
For example, evaluating the expression \lstinline[mathescape]|last [failed, True]| (where \lstinline|failed| is an expression that has no value) fails with the definition above due to the strictness of the unification operation.
In contrast, the variant based on functional patterns yields \lstinline|True| for the same expression and, thus, properly reflects the non-strict nature of Curry, which can benefit many applications in practice \cite{ChristiansenFischer2008EasyCheck,DylusEtAl2020PFLP}.

Despite their usefulness, functional patterns have the disadvantage that they must be explicitly supported by the runtime system.
For this purpose, the runtime system has to provide a special primitive for non-strict unification \cite{AntoyHanus2005FunctionalPatterns} with which functional patterns are implemented.
With this primitive, namely \lstinline|(=:<=)|, the \lstinline|last| function using functional patterns can be translated as follows.
\begin{curry}
last z | xs ++ [x] =:<= z = x where xs, x free
\end{curry}
Our original research intention was to find a way to implement functional patterns as a source-to-source transformation and, thus, without any dedicated primitive.
This way, not only the complexity of the runtime systems of different compilers could be reduced, but functional patterns could also be made available in other programming languages with similar characteristics, e.g., TOY \cite{Lopez-FraguasSanchez-Hernandez1999TOY}.

Due to the close connection between function inversion and functional patterns \cite{BrasselChristiansen2008Relation}, it seemed likely that functional patterns could also be expressed by means of function inversion.
In fact, both functional patterns and inverse functions can be expressed by one another.
For instance, we can define the inverse of \lstinline|(++)|, which non-deterministically computes all splittings of a list, using a functional pattern in the following way.
\begin{curry}
(++)§\inv§ :: [a] -> ([a], [a])
(++)§\inv§ (xs++ys) = (xs, ys)
\end{curry}
Conversely, we can transform functional patterns into calls to the inverses of the functions used in the functional patterns.
For the example of \lstinline[]|last| this transformation looks as follows.
\begin{curry}
last z = case (++)§\inv§ z of
           (xs,[x]) -> x
\end{curry}
Thus, if we had the inverse of any function at hand, we would be able to implement functional patterns directly without the need of a special primitive in the runtime system.

\section{Goal}
\label{sec:goal}

The goal of our research is to synthesize the inverse function for any given function in a lazy functional-logic programming language like Curry.
More specifically, for any function
\begin{curry}
§$f$§ :: §$\tau_1$§ -> §$\ldots$§ -> §$\tau_n$§ -> §$\tau$§
\end{curry}
we want to obtain an inverse function
\begin{curry}
§$f\inv$§ :: §$\tau$§ -> (§$\tau_1$§, §$\ldots$§, §$\tau_n$§)
\end{curry}
that complies to the following specification.
\begin{curry}
§$f\inv$§ (§$f$§ §$x_1$§ §$\ldots$§ §$x_n$§) = (§$x_1$§, §$\ldots$§, §$x_n$§)
\end{curry}
Note that we utilize functional patterns to specify the semantics of inverse functions.
In particular, inverse functions should exhibit the same non-strict behavior as functional patterns.
As a result, the implementation of \lstinline[mathescape]|(++)$\inv$| from \autoref{sec:introduction} can serve as a specification implementation against which we can test a synthesized inverse function, e.g., by using equivalence checking of non-deterministic operations \cite{AntoyHanus2018Equivalence} as implemented in CurryCheck \cite{Hanus2016CurryCheck}.

Another requirement for the synthesized inverse functions is that they should consist only of standard code.
Especially, they must not rely on functional patterns since we want to implement the latter using the synthesized inverse functions following the transformation scheme shown in \autoref{sec:introduction}.
Furthermore, using only standard code enables the application of established optimization techniques \cite{AntoyHanus2017Eliminating,Peemoeller2016Normalization}.

\section{Current Status}
\label{sec:status}

Our approach to synthesize inverse functions is based on the idea to swap the sides of rules.
We demonstrate this approach with the example of the synthesis of the inverse function \lstinline[mathescape]|(++)$\inv$|.

First, we recall the definition of \lstinline|(++)|.
\begin{curry}
(++) :: [a] -> [a] -> [a]
[]     ++ ys = ys
(x:xs) ++ ys = x : (xs ++ ys)
\end{curry}
By swapping the sides of both rules we obtain the following preliminary inverse function.
\begin{curry}
(++)§\inv§ ys           = ([], ys)
(++)§\inv§ (x:(xs++ys)) = (x:xs, ys)
\end{curry}
We can see that swapping rule sides results in new left-hand sides that may contain not only constructors and variables but also function symbols.
That is, we may get functional patterns on the new left-hand sides.
As stated in \autoref{sec:goal}, we must not use functional patterns within the definition of our synthesized inverse function.
But we can apply the transformation sketched in \autoref{sec:introduction} to eliminate the functional patterns.
We then get the following definition, which is also the final inverse.
\begin{curry}
(++)§\inv§ ys    = ([], ys)
(++)§\inv§ (x:z) = case (++)§\inv§ z of
§\phantom{\inv}§               (xs,ys) -> (x:xs, ys)
\end{curry}
Most notably, this definition complies with the specification for inverse functions from \autoref{sec:goal}.

When implementing function inversion, non-linear right-hand sides can pose a challenge.
As an example, consider the following function that appends a list to itself.
\begin{curry}
selfAppend :: [a] -> [a]
selfAppend xs = xs ++ xs
\end{curry}
If we swap the sides of a rule with a non-linear right-hand side, the resulting rule is non-left-linear which is not allowed in traditional functional-logic programs.
\begin{curry}
selfAppend§\inv§ :: [a] -> [a]
selfAppend§\inv§ z = case (++)§\inv§ z of
§\phantom{\inv}§                 (xs,xs) -> xs
\end{curry}
The intended meaning of multiple occurrences of a variable on the left-hand side is usually that the actual arguments should be structurally equal \cite{Antoy2001Constructor}.
Therefore, we can overcome the challenge of non-linear left-hand sides with the strict unification.
\begin{curry}
selfAppend§\inv§ z = case (++)§\inv§ z of
§\phantom{\inv}§                 (xs,ys) | xs =:= ys -> xs
\end{curry}

Another common challenge in inverse computation is the handling of variables which are not referred to on right-hand sides.
In the following definition, \lstinline|x| is such an extra variable.
\begin{curry}
tail :: [a] -> [a]
tail (x:xs) = xs
\end{curry}
Since extra variables can in principle take any value of their type when the function is called, the inverse function must return any possible value of the same type.
For this we can easily use free variables.
\begin{curry}
tail§\inv§ :: [a] -> [a]
tail§\inv§ xs = x : xs where x free
\end{curry}

\section{Open Issues}
\label{sec:issues}

The approach presented so far is a first step, but there are some unresolved issues.
Consider the following program that is a variant of an example that was also discussed in the context of functional patterns \cite{AntoyHanus2005FunctionalPatterns}.
\begin{curry}
g :: Int -> (Int, Int)     §\qquad\qquad§f :: Int -> Int
g 0 = (f j, j) where j free§\qquad\qquad§f i = 0
\end{curry}
Most significantly, the program contains both extra variables (\lstinline|i| in the definition of \lstinline|f|) and non-linear right-hand sides (there are two occurrences of \lstinline|j| in the definition of \lstinline|g|).
For this example, we are interested in the inverse of \lstinline|g|.
If we apply the first step of our transformation scheme to \lstinline|g|, i.e., we swap the rule sides, we get the following inverse.
\begin{curry}
g§\inv§ :: (Int, Int) -> Int
g§\inv§ (f j,j) = 0
\end{curry}
According to the semantics of functional patterns \cite{AntoyHanus2005FunctionalPatterns}, this definition of \lstinline[mathescape]|g$\inv$| should be equivalent to the following one.
\begin{curry}
g§\inv§ (0,j) = 0
\end{curry}
It is also the inverse of the following simpler definition of \lstinline|g|, which is equivalent to the one above.
\begin{curry}
g 0 = (0, j) where j free
\end{curry}
Normally, we would expect the inverses of two equivalent functions to also be equivalent.
This is, however, not the case for the two variants of \lstinline[mathescape]|g$\inv$|.
Further applying our transformation scheme to the first variant of \lstinline[mathescape]|g$\inv$| leads to the following definitions.
\begin{curry}
g§\inv§ (z,x) = case f§\inv§ z of       §\qquad\qquad§f§\inv§ :: Int -> Int
§\phantom{\inv}§            y | x =:= y -> 0§\phantom{\inv}\qquad\qquad§f§\inv§ 0 = x where x free
\end{curry}
Now consider the expression \lstinline[mathescape]|g$\inv$ (0, failed)|.
While the inverse function obtained from the simpler definition of \lstinline|g| correctly computes the value \lstinline|0| for that expression, evaluating it fails with the more complex variant of \lstinline[mathescape]|g$\inv$| due to the strict unification.
The problem arises solely from the aforementioned combination of extra variables and non-linearity in the program.
Unfortunately, we do not yet have a solution for this problem.
In this particular case, we could analyze the program, but in general it is undecidable at compile-time whether a variable is used or not at run-time.
Note that both properties alone do not pose a problem, only their combination does.

Another open issue is that we have not yet tackled the inversion of higher-order functions, a notorious difficult problem.
For instance, consider the infix application operator \lstinline|($)|, which is defined as follows.
\begin{curry}
($) :: a -> (a -> b) -> b
x $ f = f x
\end{curry}
Its inverse
\begin{curry}
($)§\inv§ :: b -> (a, a -> b)
\end{curry}
would have to return all value-function-pairs so that the function applied to the value would yield the inverse's input.
Since we cannot easily guess arbitrary functions, it seems reasonable to restrict our approach to first-order programs.
In fact, functional patterns are affected by this problem as well, so the same restriction would make sense for them too.
One possible (and well-integrated) way to do this is to introduce additional type class constraints \cite{HanusTeegen2020Data}.

\section{Related Work}
\label{sec:relatedwork}

There is earlier work on function inversion in the field of declarative programming.
Abramov and Glück presented an algorithm for inverse interpretation in a first-order functional programming language restricted to tail-recursion \cite{AbramovGlueck2000URA,AbramovGlueck2002URACorrectness}.
The so-called Universal Resolving Algorithm (URA) is based on the notion of perfect process trees \cite{GlueckKlimov1993PerfectProcessTree}.
Abramov, Glück and Klimov later extended the original URA to general recursion and improved efficiency as well as termination by reducing the search space \cite{AbramovAtAl2006URALazy}.
The reduction of the search space was achieved by inspecting partially produced output, i.e., utilizing the lazy evaluation of functional programming languages.
Furthermore, they discussed the introduction of an mgu-based equality operator that is very similar to the non-strict unification operator proposed earlier in \cite{AntoyHanus2005FunctionalPatterns}.
Another independent extension of the original URA was presented by Secher and S{\o}rensen.
They improved the termination of the URA, but at the cost of its soundness \cite{SecherSorensen2002Driving}.

Glück and Kawabe presented an approach to automatically derive the inverse of first-order functions \cite{GlueckKawabe2004Derivation}.
In the process, they used methods of LR parsing to eliminate non-determinism resulting from the automatic program inversion.
Since they aimed to obtain deterministic inverse programs in the end, they restricted themselves to injective functions in the first place.

Total inversion is subsumed by partial inversion where a part of the inputs may be fixed.
The partial inverse of a function then computes the remaining inputs with respect to the fixed ones and a given output.
Nishida, Sakai and Sakabe proposed a partial-inversion compiler of constructor term rewriting systems that first generates a conditional term rewriting system and then unravels it to an unconditional system \cite{NishidaEtAl2005Partial}.
Their approach handles extra variables by applying narrowing strategies, but only covers strict programming languages which again easily allows for non-linear right-hand sides.
Almendros-Jiménez and Vidal described another partial inversion technique---again for first-order functional programs---that is specialized on inductively sequential functions \cite{Almendros-JimenezEtAl2006AutomaticPartial}.
In the same paper, they also sketched extensions of their method to higher-order and laziness.

While most related work provide algorithms for (partial) function inversion, few publications cover the semantics and desired properties of function inversion.
For example, Braßel and Christiansen described a relational semantics for first-order-restricted Curry and also covered properties of function inversion in a lazy functional-logic programming language \cite{BrasselChristiansen2008Relation}.

\section{Conclusion}
\label{sec:conclusion}

We have presented our research on implementing functional patterns by deriving the inverse of a function from its original definition in the lazy functional-logic programming language Curry.
In essence, our idea is to swap the sides of rules and to subsequently eliminate all resulting functional patterns.
Because our synthesis generates only standard code, the resulting inverses can be subject to further optimizations.

So far our approach is limited to first-order functions whose rules do not involve both extra variables and non-linear right-hand sides.
However, we think that we already cover many relevant functions this way.
For all remaining cases we can temporarily resort to the specification implementation based on functional patterns.
The consequence is that the runtime system still has to provide the non-strict unification primitive for the time being.
With methods of static analysis we could shift the boundary, but the problem itself remains undecidable and, thus, would require additional effort at run-time.
Nonetheless, we think that being able to directly invert a relevant class of functions is a good starting point for future research.

\bibliographystyle{eptcs}
\bibliography{summary}

\end{document}